\documentclass{nature1}
\usepackage{color}
\usepackage[pdftex]{graphicx} 
\usepackage{subfigure} 
\usepackage{float}

\title{Direct Generation and Detection of Correlated Photons with 3.2 um Wavelength Spacing}

\author{Yong Meng Sua$^{1,2}$, Heng Fan$^{1,2}$, Amin Shahverdi$^{1,2,3}$, Jia-Yang Chen$^{1,2}$, \& Yu-Ping Huang$^{1,2}$}

\begin{document}

\maketitle

\begin{affiliations}
 \item Department of Physics and Engineering Physics, Stevens Institute of Technology, Hoboken, New Jersey 07030, USA
 \item Center for Distributed Quantum Computing, Stevens Institute of Technology, Hoboken, New Jersey 07030, USA
 \item Department of Electrical and Computer Engineering, Stevens Institute of Technology, Hoboken, NJ 07030, USA
\end{affiliations}

\begin{abstract}
Quantum correlated, highly non-degenerate photons can be used to synthesize disparate quantum nodes and link quantum processing over incompatible wavelengths, thereby constructing heterogeneous quantum systems for otherwise unattainable superior performance. Existing techniques for correlated photons have been concentrated in the visible and near-IR domains, with the photon pairs residing within one micron. Here, we demonstrate direct generation and detection of high-purity photon pairs at room temperature with 3.2 um wavelength spacing, one at 780 nm to match the rubidium D2 line, and the other at 3950 nm that falls in a transparent, low-scattering optical window for free space applications. The pairs are created via spontaneous parametric downconversion in a lithium niobate waveguide with specially designed geometry and periodic poling. The 780 nm photons are measured with a silicon avalanche photodiode, and the 3950 nm photons are measured with an upconversion photon detector using a similar waveguide, which attains 34\% internal conversion efficiency. Quantum correlation measurement yields a high coincidence-to-accidental ratio of 54, which indicates the strong correlation with the extremely non-degenerate photon pairs. Our system bridges existing quantum technology to the challenging mid-IR regime, where unprecedented applications are expected in quantum metrology and sensing, quantum communications, medical diagnostics, and so on.   

\end{abstract}
Quantum technology is shaping our future in a multitude of areas\cite{Gottesman1999390,Ladd201045,Yin1140,ekert,Nagata726,Rozemaprl}. Thus far, a rich variety of quantum devices and systems have been developed using disparate materials and structures, which range from trapped atoms\cite{Blatt2012} and defects in solid states\cite{Dolde2013}, to superconductors\cite{PhysRevLett.112.170501} and nonlinear nanostructures\cite{PhysRevLett.113.103601,Guo2017}, with each uniquely advantageous for certain quantum processing tasks\cite{Kurizki2015}. To access the full potential of the quantum technology, disparate quantum elements need to be synthesized to form complex, heterogeneous systems, for integrating each's distinct and complementary utility to achieve quantum supremacy. To this end, quantum correlated photons at vastly non-degenerate wavelengths are desirable to interconnect those elements and cross-link quantum processing in incompatible spectral domains, for example, by having each of the correlated photons interfering with a disparate quantum node\cite{Munroquantnode,Kimble2008,Kurizki2015,Northup2014}. Till now, a wealth of techniques have been developed for and based on quantum correlated photons, but only in the visible and near-IR spectra\cite{chocoldatomprl,Zhang2011,Allgaier2017,AleSeri2017PRX,Clausen2011}. The development in longer wavelength regimes has been scarce, which has prevented those powerful quantum techniques from spreading to the mid-IR regime (3 to 8 microns) where transformative applications in imaging, sensing, and communications reside. Only very recently, photon pairs have been generated at 3.1 microns, albeit with similar wavelengths and rather limited correlation\cite{Mancinelli2017NCTwinMIR}. 

Here we report on direct generation and measurement of correlated photons spaced by as much as 3.2 um in wavelength, between 780 nm and 3950 nm, using only room temperature devices. Quantum correlation over such distant spectra can bridge the existing quantum photonic technology in visible and near-IR into the rarely-visited, challenging mid-IR regime, thereby enabling new domains of quantum operations. As an immediate application, mid-IR single photons can be generated via heralding, by detecting the 780 nm photons with commercial, highly-efficient silicon avalanche photodiode. This photon source will give rise to a number of powerful single-photon technology in mid-IR for new capabilities in gas molecules probing, environmental monitoring, and imaging\cite{Hogstedt:14,Weibring:03,Kirmani58}.
In another application, the quantum correlation between visible and mid-IR photons can lead to exotic nonlinear interference for highly-sensitive measurements of gas molecules. This effect has been recently demonstrated for carbon dioxide molecules, whose refractive index and absorption in mid-IR are determined by using solely visible optics\cite{Kalashnikov2016}.The third application may be found in free-space quantum communications, for which light around 4 micron has been measured to experience much less scattering and scintillation comparing with other commonly used wavelengths\cite{Zhang:17}. Falling in a transparent window of the atmosphere with an acceptable background level of blackbody radiation, the 3950 nm photons are superior for free-space quantum communications under dynamic, scattering weather conditions, such as in urban or maritime environment. With their partner photons at 780 nm---a convenient wavelength for quantum memory\cite{GCGuoNCRBEITQuantMemo} and detection with silicon avalanche photodiodes (Si-APD)---several quantum-communication protocols are ready to be implemented, including the Ekert protocol\cite{ekert}, a high-dimension version of BB84 protocol\cite{Lima16DQKDSCIREP},quantum repeaters protocol\cite{ChrisSimonQuantRepPhoPairPRL2007}, and quantum secure direct communication protocol\cite{PhysRevA.68.042317,PhysRevLett.118.220501}.  

Our photon pairs are created through spontaneous parametric down conversion (SPDC) in a magnesium-doped periodically-poled lithium niobate (MgO:PPLN) waveguide. Although similar waveguides have been used for SPDC in the visible and near-IR regimes, here the extreme non-degeneracy between the photons poses significant challenges in the spatial-mode and phase matching in the waveguide. In response, our waveguide's geometry is carefully tailored to support overlapping fundamental transverse modes for all lightwaves, and its optical domains are periodically poled to offset the large phase mismatch. By fine tuning and stabilizing the waveguide's temperature, a high -87dB SPDC efficiency is achieved. 

The generated 780 photons are detected directly by a free-running Si-APD with ultralow dark count ($<1.4$ counts/second). However, there is no available photon detector for the mid-IR photons at room temperature. The existing mid-IR detectors require cryogenic cooling and suffer low detection efficiency yet high dark count\cite{Rajan2012SSPD,Rath2015}. Instead, our 3950 nm photons are measured by an upconversion photon detector, consisting of another PPLN waveguide to convert them to 651 nm for detection in a second Si-APD. The waveguide, which has of similar geometry and periodic poling with the SPDC waveguide, achieves 34\% internal conversion efficiency with 2.6$\times 10^{-6}$ photon noise per time-frequency mode. Meanwhile, it acts a filter to reject background photons not falling in the phase matching band for the conversion. Furthermore, by using broadband, mode-shaped pumping pulses, it can also selectively upconvert the signal photons over interfering noises that overlap in both time and spectrum\cite{Shahverdi2017}. These unique capabilities offered by such upconversion detectors will prove crucial for photon-starving applications, such as free-space quantum communications over free space and remote sensing, where the received signal is swamped by ambient photons.     

The correlation measurement between the 780 nm and 3950 nm photons yields a coincidence to accidental ratio of 54, which indicates high quantum-state purity of the pairs. This gives a Franson visibility of 96\% in the context of time-energy entanglement\cite{Ma:09}, which is on par with existing SPDC sources in visible and near-IR. It also suggests that, by detecting the 780 nm photons, mid-IR photons can be generated via heralding with  $g^{2}(0)=0.08$, making it a high-quality single photon source in mid-IR. 

The remaining of this paper is organized as follows. In Section 1, we describe the experimental setup. In Section 2 and 3, we present the results on single-photon detection and photon-correlation measurement, respectively. In Section 4, we present a conclusion and prospective.  

\section{Quantum correlated photon pair generation.}
\begin{figure}
\includegraphics[width=6in]{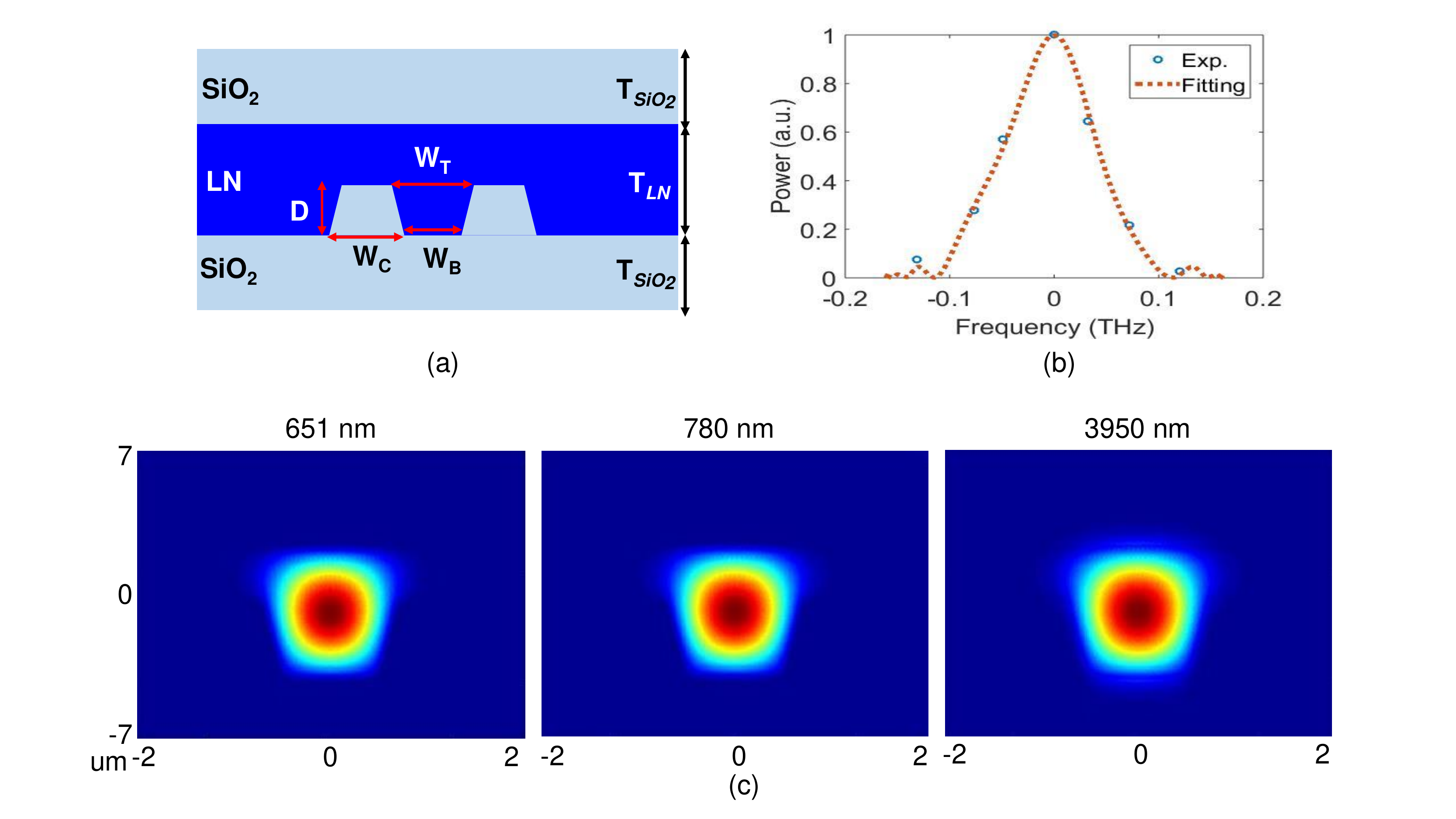} 
  \centering
\caption{(a) Cross-sectional view of the SPDC and upconversion PPLN waveguide (HC Photonics Corp.), with dimensions (in micron): D=3.5, $T_{LN}$=6.2, $W_B$=7.8,  $W_T$=11, $W_C$=16, $T_{SiO2}$=0.7 (b) The measured phase matching profile of the upconversion waveguide (the PSDC waveguide is similar). Blue dots show the experiment results and dashed line is the simulation result of the upconverted optical power plotted against the frequency offset of upconversion pump wavelength (c) The simulated fundamental quasi-TM modes of SPDC and upconversion waveguides for the pump (651.2 nm), signal (779.8 nm) and idler (3950 nm) wavelengths for the waveguide geometry in (a).}
\label{fig1}
\end{figure}

SPDC, a quantum nonlinear process in which a pump photon is converted to a pair of signal and idler photons, has been used ubiquitously for photon-pair generation, mostly in the visible and near-IR spectra\cite{kwiat1999high,Mancinelli2017NCTwinMIR}. It is efficient only when the pump, signal, and idler satisfy the energy and momentum conservation: 
\begin{eqnarray}
k_p - k_s - k_i +\frac{2\pi}\Lambda = 0, \\
\omega_p -\omega_s -\omega_i  = 0, 
\end{eqnarray}
where $k_j$ and $\omega_j$ ($j=p,s,i$) are the wave number and angular frequency, respectively, for the pump, signal, and idler, and $\Lambda$ is the periodic poling period. For the current SPDC process where a 651.2 nm pump photon downconverts to a pair of a 779.8 nm signal photon and a 3950 nm idler photon. The detuning of the pair is over 3.2 microns, or $>$300 THz, which is the largest detuning thus far in any waveguide-based photon source. Because of this extreme nondegeneracy, there are two major challenges in attaining efficient SPDC. First, the transverse modes for the signal and idler are very dissimilar, which can cause significant transverse-mode mismatch thus a low nonlinear coupling efficiency. To overcome this difficulty, the waveguides are designed to have dimensions comparable to the idler wavelength, i.e. a silicon-dioxide cladded core with 3.5 um height (D) and  11 um top width ($W_t$) ridge structure as shown in Fig.~\ref{fig1} (a), so that all lightwaves are in their respective fundamental quasi-transverse-magnetic (quasi-TM) modes for the maximum overlap.  Figure~\ref{fig1} (c) shows quasi-TM modes for the pump, signal and idler wavelengths, which are simulated via finite elements method (FEM) based on the parameters extracted from our actual waveguide, indicating a good mode overlap. Second, there is a large group velocity difference between the signal and idler photons, leading to strong phase mismatching. In response, the waveguide is periodically poled with a domain length of 16.7 micron to achieve quasi-phase matching for those fundamental transverse modes. The temperature of the PPLN waveguide is fine tuned and stabilized at 32.2 $\pm$ 0.1 $^\circ$C to maintain the optimum phase matching for the highest SPDC efficiency. 

The phase matching properties for the SPDC and upconversion waveguides are nearly identical, as they are constructed similarly. To measure their bandwidth, we use weak coherent pulses centered at 3950 nm, created by difference frequency generation between strong 651.2~nm pulses and seeding pulses at 779.8~nm. The same 3950 nm pulses are used to fine tune the waveguide's temperature and optimize the optical beam alignment for the maximum conversion for the idler photons. The phase matching of the upconversion waveguide is then characterized by scanning the wavelength of the upconversion pump while fixing the weak coherent pulses, both maintained at constant power. The upconverted light is measured by using a high sensitivity optical spectrum analyzer (OSA). The measured phase matching results are plotted in Fig.~\ref{fig1} (b), where a clear phase-matching peak is shown.  
%


\begin{figure}
\includegraphics[width=6in]{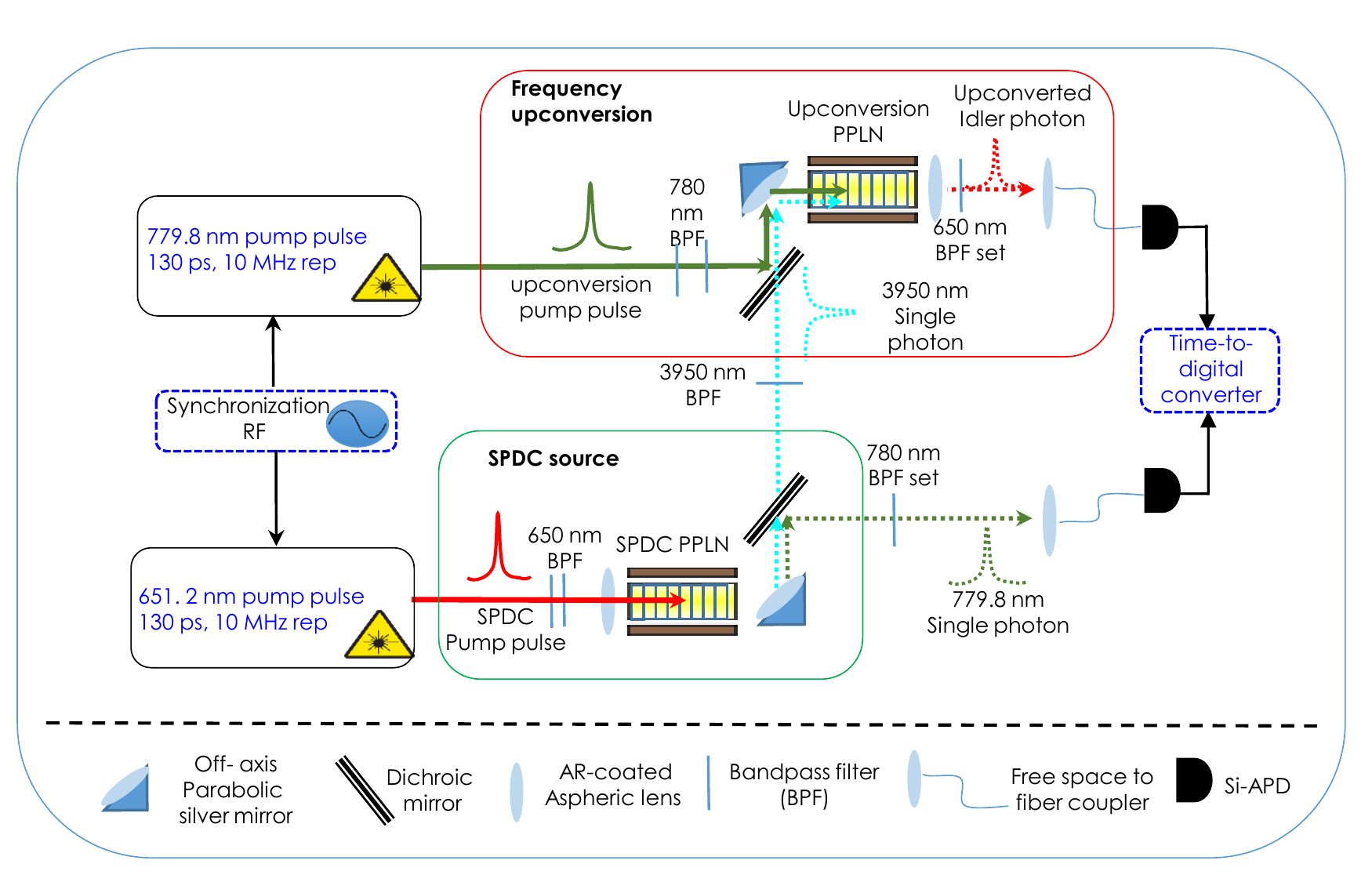} 
  \centering
\caption{\footnotesize Experimental setup: A 651.2 nm pulse train with 10 MHz repetition rate and 130 ps pulse width is coupled into a 2.2-cm long PPLN waveguide through an anti-reflection coated aspheric lens. The waveguide's temperature is stabilized at 32.2~$\pm$~0.1 $^\circ$C using a temperature controller (not shown). The output photon pairs at 779.8 nm and 3950 nm are collimated by using an off-axis parabolic mirror and separated into two paths at a dichroic mirror. The signal photons are first guided through a 780 nm bandpass filter set in free space (bandwidth: 3 nm; extinction: 140 dB) which rejects the pump pulses while defining the detection bandwidth, before collected by a fiber collimator and detected by a free-running Si-APD. The idler photons are steered through a 3950 nm bandpass filter (bandwidth: 90 nm; extinction: 70 dB) and combined with a 779.8 nm pulse train with 130 ps pulse width and 10 MHz repetition rate that is synchronized with the SPDC pump. The two are then focused by using another off-axis parabolic mirror and coupled into a second PPLN waveguide for frequency upconversion of the idler photons to 651.2 nm. The upconversion waveguide is similar to the SPDC wavegude but with temperature stabilized at 62.2~$\pm$~0.1 $^\circ$C for the optimum phase matching. The upconverted photons are picked by a 650 nm bandpass filter set that rejects the pump by more than 160 dB, as well as any out-of-band background photons created by in the waveguide. Those photons are then coupled into a fiber collimator and detected by another free-running Si-APD. The coincident detections between the two Si-APD are recorded with a time-to-digital-converter (TDC).}
\label{fig2}
\end{figure}

Figure \ref{fig2} shows a schematic of the experimental setup, where the photons are created in a PPLN waveguide through SPDC: 651.2~nm $\rightarrow$ 779.8~nm + 3950~nm. The signal photons are filtered and detected by using a free-running fiber coupled Si-APD. Here the Raman scattering photons of the pump pulses falling in the signal band is not expected to be significant due to the large detuning between the pump and signal wavelength of more than $2500~cm^{-1}$ \cite{pelc2011long}. The idler photons are first frequency upconverted to 651.2 nm using another PPLN waveguide pumped by 779.8~nm pulses (779.8~nm +~3950~nm $\rightarrow$~651.2~nm) and eventually detected by a second Si-APD. More details are given in the Method section.

\section{Single Photon Detection}

\begin{figure}[H]
  \centering
   \subfigure[]{
   \label{fig:subfig1.1.1:a} 
    \includegraphics[width=3.2in]{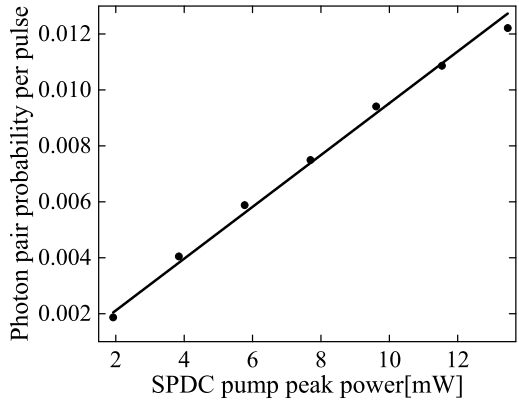}} 
  \subfigure[]{
   \label{fig:subfig1.1.1:b} 
    \includegraphics[width=3.15in]{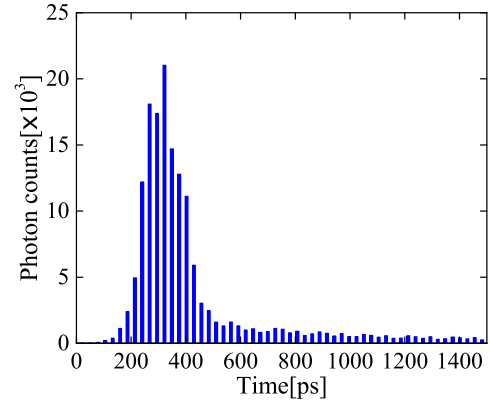}}
   \caption{(a) Measured signal photon generation probability as a function of the SPDC pump peak power, where black dots are the experimental data and the black solid line is a linear fit. The error bars, 1 s.d. (are smaller than the symbol size) assuming Poissonian counting statistics. (b) Histogram of the signal photon arrival time by the means of time-correlated single photon counting measurement integrated in 100 seconds.}
  \label{fig3} 
\end{figure}

At 780 nm, the signal photons are measured directly using a Si-APD after filtering. Figure~\ref{fig3} (a) plots the measured signal-photon generation probability as a function of pump peak power, which displays a linear dependence, as expected. From the slope, the SPDC efficiency is estimated to be -87 dB, meaning that about one in every $5\times10^{8}$ pump photons probalistically downconverts into a pair of correlated photons. Figure~\ref{fig3} (b) plots the histogram of the signal photon arrival time by the means of time-correlated single photon counting, shown a single-photon pulse of 130 ps full width at half maximum (FWHM) after the filters. The cascaded filters centered at 780 nm provide an effective spectral bandwidth of 3 nm (FWHM), so that there are effectively 100 detected time-frequency modes\cite{PhysRevXtimefreqmode2015}.

In contrast, direct detection of mid-IR photons has been demonstrated using cryogenically-cooled superconducting nanowire single photon detectors (SNSPD), whose responses in principle cover a wide spectrum from visible to mid-IR\cite{MArsilinanoletter5um,MarsiliF.2013}. However, there is a sharp decline in the detection efficiency accompanied by a rapid rise of intrinsic dark count in the mid-IR regime\cite{MarsiliF.2013}, which impedes its utility. To overcome this difficulty, upconversion detection has emerged as a viable alternative and as shown very recently, has an edge over SNSPD's for dark-count susceptible applications\cite{Mancinelli2017NCTwinMIR}. Here, we utilize quantum frequency conversion to upconvert the 3950 nm photons in a PPLN waveguide into a visible wavelength at 651.2 nm and subsequently detected by a low-dark count (1.4 counts per second) Si-APD at free running. The phase matching bandwidth of the waveguide is measured to be 5.6 nm (FWHM), corresponding to 107 GHz in spectral bandwidth. This rather narrow phase matching bandwidth intrinsically filters out background photons outside the idler band. Even for those in-band noise photons, further rejection is attainable by using broadband, mode-shaped upconversion pump pulses approach the phase matching limit\cite{Shahverdi2017}. This is especially advantageous for quantum applications in the mid-IR regime where substantial background noise exists, such as from the blackbody radiation.



\begin{figure}[H]
  \centering
   \subfigure[]{
   \label{fig:subfig1.1.1:a} 
    \includegraphics[width=3.2in]{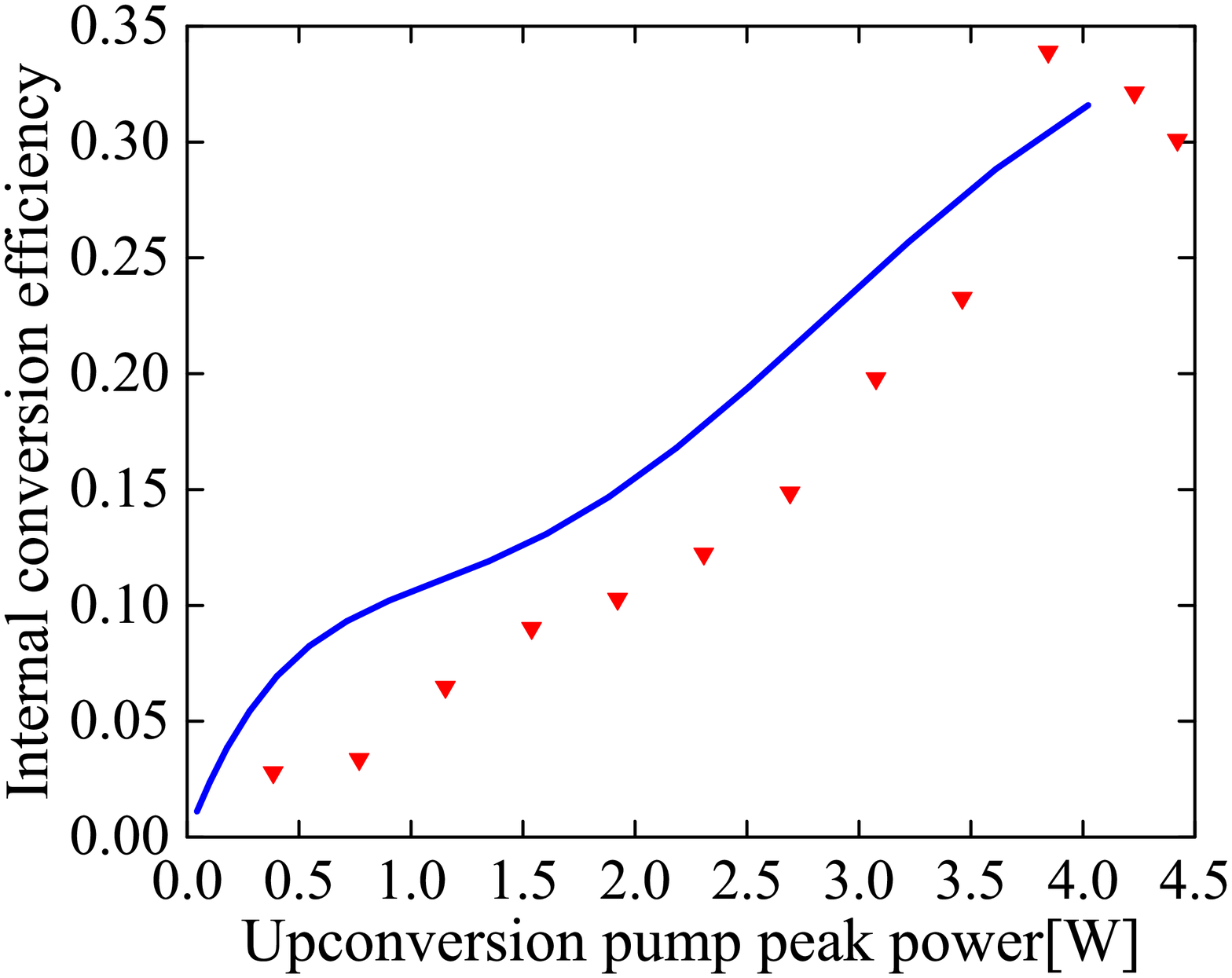}} 
  \subfigure[]{
   \label{fig:subfig1.1.1:b} 
    \includegraphics[width=3.15in]{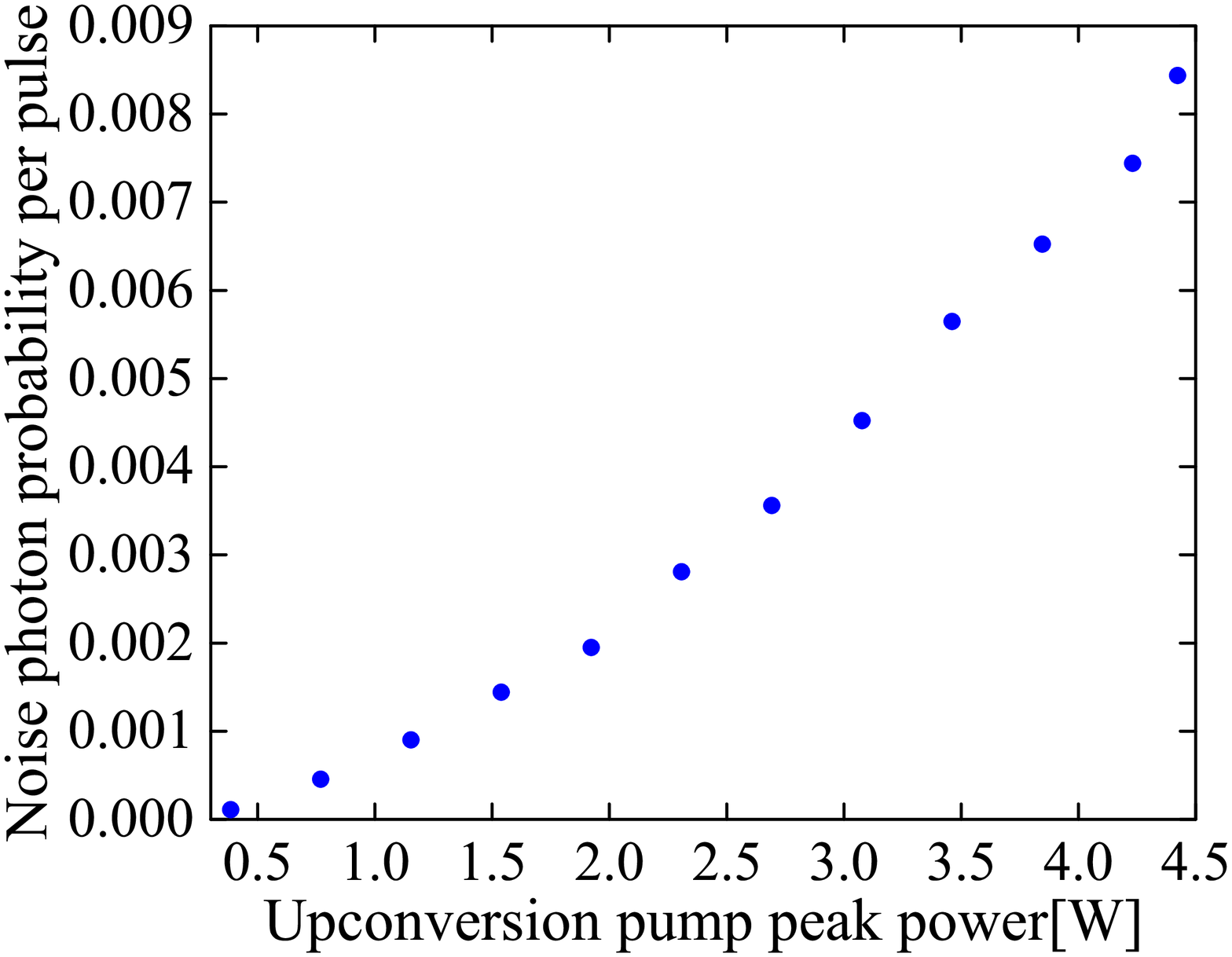}}
  \caption{(a) The internal conversion efficiency of the upconversion waveguide. Red triangles are the experiment data and the blue solid line is the numerical simulation without any free fitting parameter. (b) The noise photon rate created by the upconversion pump as a function of its peak power. The error bars are smaller than the symbol size.} \label{fig4} 
\end{figure}

To contrast the upconverted idler photons with the Raman-scattered photons, we record the total single photon counts with the SPDC on and off, from which the true-upconverted idler-photon counts are derived through subtraction of the latter from the former. 
Then, the internal upconversion efficiency was characterized by dividing the true upconverted idler-photon counts by the correlated signal-photon counts at 780 nm over the same integration period (see the following section), after accounting for all linear losses and the quantum efficiency of each Si-APD, as explained in detail in the Method section. The result as a function of the upconversion pump peak power is shown in Fig.~\ref{fig4}(a), with the highest conversion efficiency $33.9\pm1.8\%$ achieved at the peak pump power of 3.8 W. The efficiency decreases as the power further increases. This saturation of conversion efficiency is likely due to the existence of multiple spatial modes for the photons.  

To understand this result, we numerically simulate the frequency upconversion of the idler photons at 3950 nm by strong pump pulses at 779.8 nm. First, the normal modes of the photon pairs and the photon pair generation probability for each mode  are calculated via the modal decomposition of their joint spectral function---determined by the pump profile and phase matching for SPDC, and the filter profile for the signal photons \cite{YPHuang2010PRA}. This calculation finds a total number of SPDC modes to be about 15, approximately equal to BT/4, with B the phase matching bandwidth of the SPDC waveguide and T the temporal width of the SPDC pump pulses. Then, the upconversion process for the idler photons is modeled with a set of coupled Heisenberg equations, which are numerically solved using a split-step Fourier transform method \cite{Kowligy:14}. The conversion efficiency for each SPDC mode is simulated with the measured temporal profile of the upconversion pump pulses and the upconversion waveguide's phase matching profile. The simulation results for different peak power of the upconversion pump, which assumes no free fitting parameter, is plotted in Fig.~\ref{fig4}(a). It shows a good agreement with experimental results, with the slight discrepancy likely due to the existence of higher-order spatial modes for the pump pulses and the non-negligible propagation loss in the waveguide, both of which have been neglected in the present simulation.

Figure \ref{fig4}(b) shows the number of noise photons created by the upconversion pump. With a 33.9\% conversion efficiency, the noise photons are approximately 6.5$\times 10^{-3}$ per pump pulse for the 144 detected modes entering the Si-APD (see Fig.~4(b)). This can be lowered substantially to 4.5x10$^{-5}$ per pulse if the number of the detected modes is reduced to 1, by either using shorter pulses for the pump and SPDC photons, or using narrower bandpass filters for the upconverted photons. The amount of noise photon counts is dependent on the number of the detected time-frequency modes, as we show in Method section. Thus, it is crucial to reduce it, in order to improve the signal-to-noise ratio which will in turn minimizing key error rate for real world QKD. Also, if the conversion efficiency can be further increased, less pump power will required, so that the Raman photons will be suppressed. Taking into account the coupling losses, upconversion efficiency (~34$\%$), and quantum efficiency (22\%) of the Si-APD, the total detection efficiency of the 3950~nm single photons is approximately 0.078$\%$ for the current under-optimized setup. It can be improved significantly to 2.5\% by replacing the present parabolic mirrors by ones with a shorter focal length to reduce the total waveguide coupling loss by 6 dB, and using highly efficient Si-APD's with over 80\% quantum efficiency.       


\section{Coincidence measurement of quantum correlated photon pair}
\begin{figure}[H]
  \centering
  \subfigure[]{
   \label{fig:subfig1.1.1:a} 
    \includegraphics[width=3.2in]{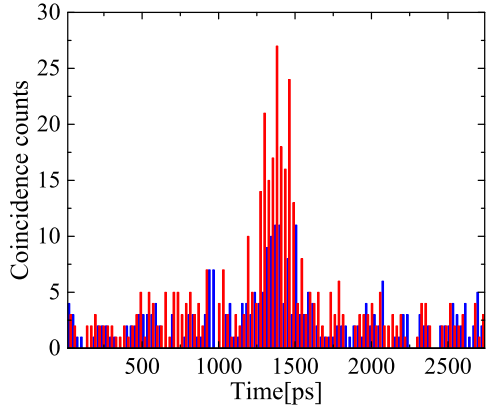}}
  \subfigure[]{
   \label{fig:subfig1.1.1:b} 
    \includegraphics[width=3.2in]{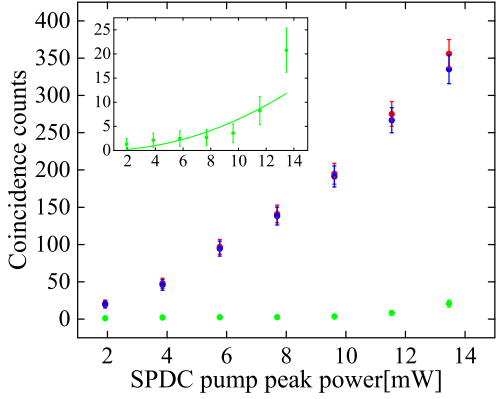}}
  \subfigure[]{
   \label{fig:subfig1.1.1:b} 
    \includegraphics[width=3.15in]{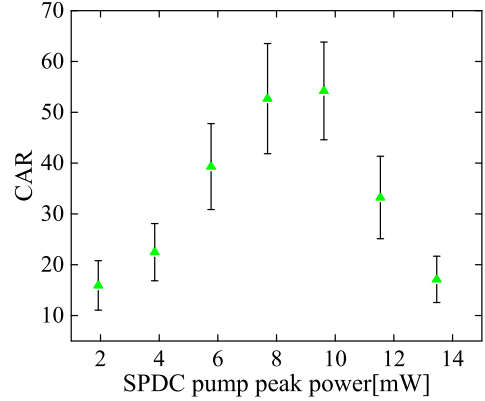}}
  \caption{(a) Histogram of the raw coincidence counts obtained with (red) and without (blue) input of 3950 nm idler photons. Contribution from the upconversion noise photons and detector dark counts are not subtracted. (b) Measured coincidence (red square), accidental coincidence (green dot), true coincidence (blue dot), plotted as a function of pump peak power. (Inset) accidental-coincidence(green dot) plotted on a smaller scale with quadratic fitting (green solid line). (c) True coincidence-to-accidental coincidence ratio versus pump peak power. Each data point is acquired over an integration time of 300 seconds. The error bars are calculated assuming Poissonian photon counting statistics.}
  \label{fig5} 
\end{figure}

To quantify the purity of the quantum-correlated photon pairs, we perform photon coincidence measurement between the signal (780 nm) and idler (3950 nm) photons. A coincident detection is registered when both Si-APD's click within the coincidence window. In contrast, an accidental-coincidence detection is recorded when both Si-APDs click outside the coincidence window. Figure \ref{fig5}(a) plots a typical coincidence histogram between two Si-APDs, which exhibits an obvious peak within the coincidence window, illustrating a pronounced timing correlation between the signal and idler photons. We measure the total coincidence and accidental-coincidence as a function of the pump peak power. The true-coincidence counts---which result from the simultaneous detection of the photon pairs---are obtained by subtracting the accidental coincidence from the total coincidence. As shown in Fig.~\ref{fig5} (b), the true coincidence is linearly dependent on the pump power, as expected for SPDC. In contrast, the accidental-coincidence arises from simultaneous detection of two uncorrelated photons, and thus depends quadratically on the pump peak power, as shown in the Inset of Fig.~\ref{fig5} (b). 

The true coincidence to accidental-coincidence ratio, CAR, is plotted in Fig.~\ref{fig5}(c) as a function of the SPDC pump peak power. A maximum CAR of 54$\pm 7$ is obtained at a peak power of 9.7 mW, limited by the uncorrelated noise photons and dark counts of the upconversion detection system. It can be improved by reducing the number of detected time-frequency modes of the signal and idler photons to 1\cite{YupingPRA2011}. For lower pump power, the CAR is only around 10$\sim$20, which attributes to the low photon-pair generation rate relative to the total dark count rate of the entire detection system. At high pump power, on the other hand, the multiphoton effect involving the simultaneous generation of more than a photon pair increases the probability of accidental-coincidence detection, which leads to lower CAR values\cite{Arahira:11}. At the highest CAR of 54 $\pm 7$, we obtained a true coincidence count of 275 within an integration time of 300 s. Considering the total detection efficiency of 1.3$\%$ and 0.078$\%$ for the signal and idler, respectively, the photon pair production rate is about $9.4\times 10^{4}$ pair per second, with spectral brightness of $4.4\times 10^{2}$ pair/s/nm/mW. We expect to improve the production rate and spectral brightness by two orders of magnitude by simply using GHz-repetition-rate pump pulses. 

With CAR of 54$\pm 7$, this SPDC source can already be used to generate two photon time-energy entangled states with Franson-inteferometry visibility $\frac {CAR-1}{CAR+1}= 96\pm 0.6 \%$\cite{Ma:09}, on par with SPDC sources in common wavelengths\cite{Sarrafi:14,Autebert:16}. Combined with appropriate quantum error correction or privacy amplification, our SPDC scheme can be readily applicable for free-space QKD based on time-energy entanglement in the mid-infrared spectrum. On the other hand, by adopting idler photon at 3950 nm for free-space quantum channel and storing its twin photon at 780 nm photon with a rubidium-based quantum memory, this SPDC source is an ideal candidate for quantum secure direct communication protocol with quantum memory\cite{PhysRevLett.118.220501}. Furthermore, by detecting the signal photons at 780 nm, the generation of idlers photons at 3950 nm can be heralded, with a second-order correlation $g^{2}(0)=0.08 \pm 0.015$\cite{LixiaoyingPhysRevA.83.053843}. By substituting the PPLN waveguide with a PPLN crystal, the same strong correlation of the photon pairs can be translated to the spatial domain for heralded or quantum ghost imaging applications in mid-IR, in which less than one photon per pixel is sufficient to image an object\cite{Morris2015NCheraldimage}.

\section{Discussion/summary}
We have demonstrated direct generation and detection of high-purity photon pairs across the visible/near-IR (780 nm) and mid-IR (3950 nm) spectra, using only room-temperature optics. The generation utilizes spontaneous parametric downconversion in a lithium-niobate waveguide with specially designed geometry and periodic poling. The detection of 3950 nm uses an upconversion photon detector consisting of another similar lithium-niobate waveguide for quantum frequency conversion, followed by a silicon avalanche photodiode.

Our results highlight a viable approach to heterogeneous quantum systems by integrating incompatible quantum devices and processes using highly-nondegenerate, quantum correlated photons. This may prove to be crucial for achieving quantum supremacy, as there is no single material platform capable of hosting all necessary quantum functionalities.  Rather, a breadth of disparate quantum devices have exhibited practical advantages for certain, individual quantum tasks. By efficiently integrating them, scalable, practically-connected quantum systems are possible to establish practical advantages. 

For the near-term applications, our technique can be adopted quickly for quantum enhanced technology of sensing and imaging in mid-IR by leveraging the extensive developments in the visible/near-IR regime. Prospective applications includes single photon microscopy and quantum-enhanced spectroscopy for sensitive biomedical imaging and diagnostics. It also unlocks a new paradigm of weather-proof quantum communications using 3950 nm photons, which promises to mitigate the atmosphere disturbance and extend the reaching range. 

The immediate future work is to adopt the quantum optical arbitrary waveform technique for both photon pair generation and upconversion detection, which will potentially improve the coincidence-to-accidental coincidence ratio by $>$20 times (to over 1000) while further suppressing background noise photons. The same technique can be used to create high-dimensional entanglement, which will further increase the robustness of the quantum free space applications against even some of the most challenging atmospheric conditions.

\begin{methods}
\subsection{Experimental setup for photon-pair generation}
In the experimental setup shown in Fig.~\ref{fig2}, the 10 MHz-repetition rate, 130 ps pump pulses at 651.2~nm is generated via sum frequency generation between amplified optical pulses at 1555.1 ~nm and a continuous-wave laser at 1120.4~nm, through 1555.1~nm + 1120.4~nm $\rightarrow$~651.2~nm. Both the 1555.1~nm and 1120.4~nm beams are combined by using a dichroic mirror before focused and coupled into a 25-cm long MgO:PPLN waveguide using an aspheric lens. The lens is chosen to match the mode field diameter and numerical aperture (NA) of the waveguide for the optimum coupling, with efficiencies at 35$\%$ and 30$\%$ for each. The output pulses at 651.2~nm are collimated using another high-NA ashperic lens. A set of cascaded bandpass filters centered at 650 nm are used to clean up the residual 1555.1~nm and 1120.4~nm light with $>$130 dB extinction ratio, which also rejects most Raman scattering photons in the waveguide. Then, a sequence of a quarter-wave plate, a half-wave plate, and a polarizing beam splitter is used to rotate the polarization state of the pump pulses into the vertical polarization, as required for the SPDC. 

The PPLN waveguide for SPDC is enclosed in a metal-block housing, whose temperature is stabilized at 32.2~$\pm$~0.1 $^\circ$C. The output of SPDC photons are collimated by using a 90-degree off-axis parabolic silver mirror with a focal length of 12.5 mm in order to achieve achromatic collimation and good coupling for the largely nondegenerate signal and idler photons. Separated by a dichroic mirror, the signal photons at 780 nm are guided through cascaded bandpass filters centered at 780 nm with 3 nm FWHM-bandwidth, to reject the residual pump pulses and the out-of-band photons from spontaneous Raman scattering by more than 140 dB extinction. Then, the single photons are coupled into a single-mode fiber (SM600) and detected by a free-running, fiber-coupled Si-APD (ID100, ID Quantique) with 12.5 $\%$ quantum efficiency and low dark-count  rate ($<$1.4 Hz). The 3950 nm photons are picked by an infrared narrow band filter (center wavelength: 3950 nm,  FWHM bandwidth: 60 nm) and upconverted to be detected by a Si-APD (ID100, ID Quantique) with 22.5$\%$ quantum efficiency and low dark count ($<1.8$ Hz). A detected photon produces a TTL pulse, which is fed into a multi-channel time-to-digital converter (SENLS, HRM-TDC) for coincidence measurement and analyses. The Si-APD has 50 ps timing resolution while the TDC has 66 ps root-mean-square timing resolution and 27 ps time-bin resolution.

\subsection{Quantum frequency upconversion}
The strong pump pulses for upconversion are generated via second-harmonic generation of amplified 1559.6 nm pulses in a 10-mm-long PPLN waveguide to generate 779.8 nm pulses with 10-MHz repetition rate, 130 ps FWHM pulse width. Using a dichroic mirror, both idler photon and the upconversion pump pulse are focused and coupled into the PPLN waveguide via a 90 degree off-axis parabolic silver mirror (FL=12.5 mm). The coupling efficiency is about 10\% for the idler photons and 25\% for the upconversion pump pulses. A set of cascaded optical bandpass filters (center wavelength: 650 nm; FWHM bandwidth: 3 nm) are then used to reject the residual pump pulses and out-of-band spontaneous Raman scattering photons by more than 160 dB of extinction. 

To determine the internal conversion efficiency of the upconversion waveguide, we first measure the photon-pair generation rate by measuring the signal single photons at 779.8 nm. By using narrow bandwidth bandpass filters with high extinction ratio, we make sure the detections acquired by the Si-APD are mainly from SPDC with the signal-to-noise ratio (SNR) by more than 30 dB. The center wavelength of the narrow bandpass filters for the 3950 nm photons and the subsequent upconverted 651.2 nm photons were chosen accurately in accordance with the phase matching conditions. By accounting for the independently characterized total loss from the output of the SPDC waveguide to the Si-APD, and the quantum efficiency of the Si-APD at the signal wavelength (~12.5$\%$), we calculate the photon pair generation rate. Then, we can deduce the internal conversion efficiency of the upconversion waveguide after including the total transmission loss before reaching Si-APD (19 dB) and the quantum efficiency of the Si-APD at 650 nm (~22$\%$). The total loss from the output of the SPDC waveguide to the output of the upconversion waveguide is measured to be about 16 dB by using a weak coherent light at 3950 nm generated by difference frequency generation (DFG) from the same SPDC waveguide. About 3 dB loss is caused by the output coupling loss of the off-axis parabolic mirror, 1 dB of loss is from the 3950 nm bandpass filters and dichroic beam splitters, and another 12 dB is from the input coupling efficiency and propagation loss of the upconversion waveguide. In our setup, the coupling efficiency is mainly limited by the off-axis parabolic mirror. 

The input coupling loss for the upconversion waveguide can be reduced by using anti-reflection-coated aspherics lens and optimizing the coupling with individually shaped spatial mode of upconversion pump and idler photons. The optical loss endured by the upconverted 650 nm photons due to the off-axis parabolic silver mirror, bandpass filters, free space to fiber coupling is about 4 dB before reaching the Si-APD. 

\subsection{Noise photon counts for different time-frequency modes}
We measure the noise photons without the input of the idler photons with approximately 144, 725 and 2900 detected time-frequency modes for different pump peak power, as shown in Fig.~\ref{fig6} (a) and (b). As the intrinsic dark count and after pulsing probability of the Si-APD are negligible, the majority of the dark counts are from the spontaneous Raman scattering of the upconversion pump pulses that falls in the window of the bandpass filter and subsequently detected by the Si-APD. As shown, the noise count rate is higher with wider bandpass filter (which means more detection modes) and is linearly dependent on the pump peak power. This highlights the importance of using a narrow bandpass filter matching with upconverted idler photons for achieving optimum signal-to-noise ratio. The noise detection at optimum conversion efficiency is about 3.8$\times 10^{-4}$ per pulse with 144 detection modes, which includes the detector dark count, noise photon due to frequency upconversion, blackbody radiation from the environment and inside the waveguides. Several approaches can be implemented to further suppress the noise count probability. For example, we can employ the quantum optical arbitrary waveform technique to tailor the upconversion pump pulse in accordance with the phase matching bandwidth to maximize the noise rejection of the blackbody radiated noise photons in mid-IR. Additionally, we can minimize the noise photon due to Raman scattering of pump pulse by using a much more efficient upconversion PPLN waveguide which will substantially lower the requirement of the peak power\cite{HepingZeng2013APL}. The rejection of the pump-induced Raman photons can be further improved by using a volume bragg grating filter which can provide a much narrower bandwidth yet with little transmission loss for the upconverted photon\cite{Kuo:13}.

\begin{figure}[H]
  \centering
   \subfigure[]{
   \label{fig:subfig1.1.1:a} 
    \includegraphics[width=3in]{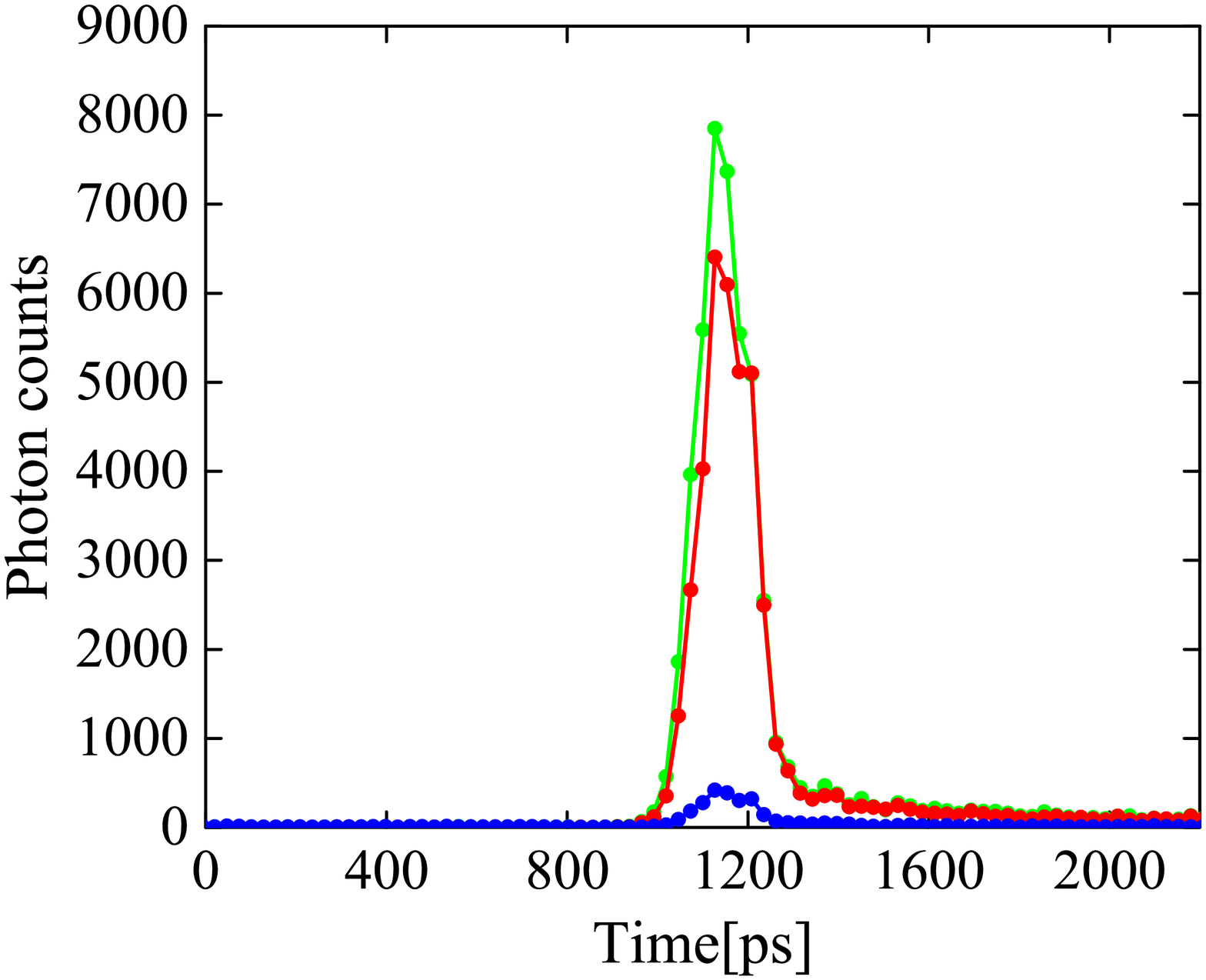}} 
  \subfigure[]{
   \label{fig:subfig1.1.1:b} 
    \includegraphics[width=3in]{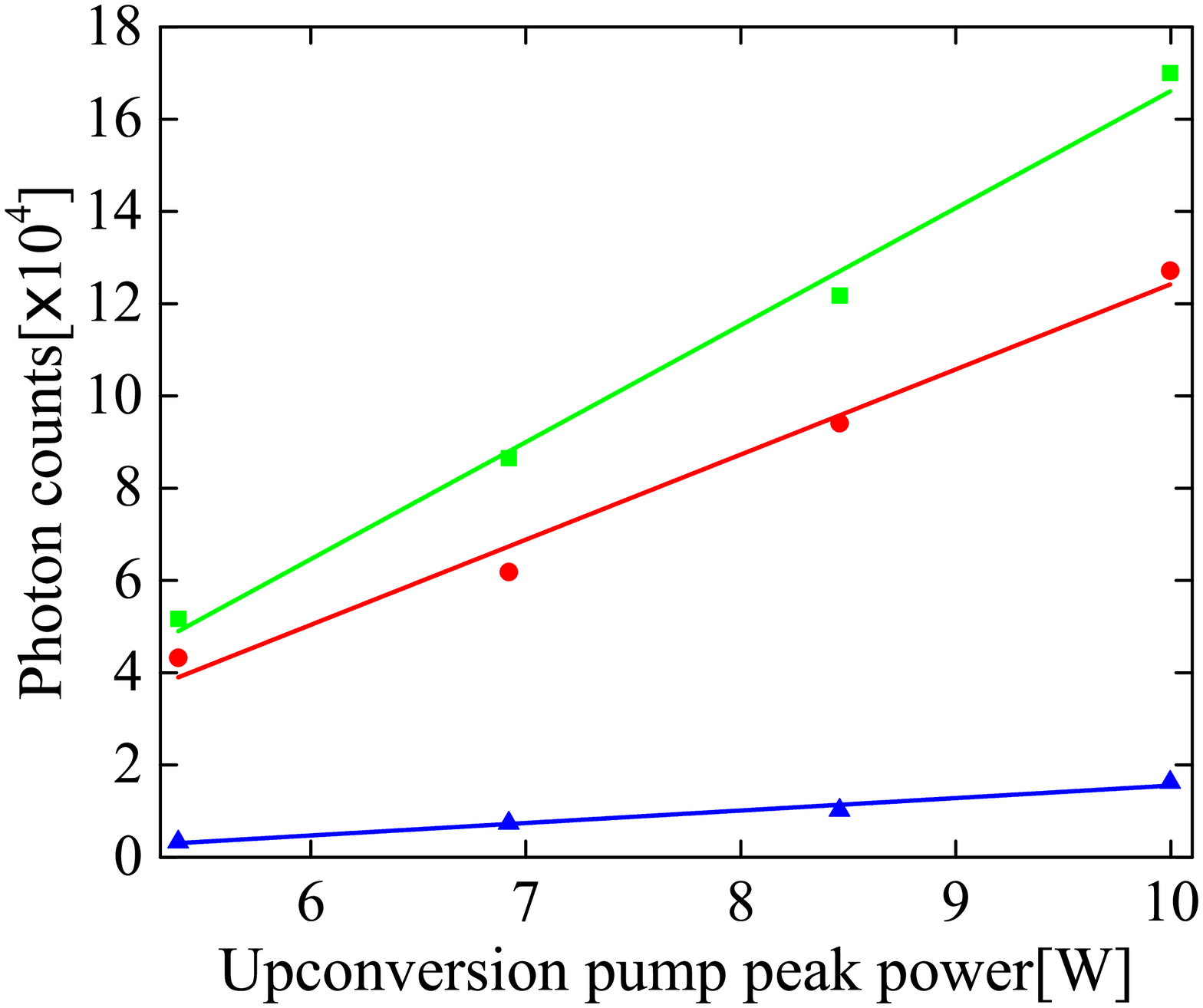}}

  \caption{(a) Histograms of the noise photon detection for 144 (green), 724 (red) and 2900 (blue) time-frequency modes over 100 second integration time. (b) Single photon counts obtained over 100 second integration time for the 144 (green), 724 (red) and 2900 (blue) time-frequency modes versus the pump power. The error bars assume Poissonian counting statistics and are smaller than the symbol size.} 
  \label{fig6} 
\end{figure}
\subsection{Quantum frequency up-conversion scheme}



\end{methods}



\begin{addendum}
 \item This research was support in part by the Office of Naval Research (Award No. N00014-15-1-2393) and by the National
Science Foundation (Award No. ECCS-1521424). We appreciate Prof. Rainer Martini's enlightening discussions on  experiment design. We also acknowledge HC Photonics for PPLN waveguide design, fabrication and extensive technical support.

\item[Contribution] Y.M.S. and F.H. designed and performed the experiment. Y.M.S., F.H. and Y.-P.H. analyzed and interpret the data. A.S. and J.Y.C. performed the numerical simulation. Y.-P.H. conceived and supervised the project. All contributed to manuscript preparation. 

 \item[Competing Interests] The authors declare that they have no
competing financial interests.
 \item[Correspondence] Correspondence and requests for materials
should be addressed to yuping huang ~(email: yuping.huang@stevens.edu).
\end{addendum}



\end{document}